# Measuring Network Robustness by Average Network Flow

Weisheng Si [1], Balume Mburano [1], Wei Xing Zheng [1], and Tie Qiu [2]

[1] School of Computer, Data and Mathematical Sciences, Western Sydney University, Australia

[2] School of Computer Science and Technology, Tianjin University, China

{w.si, b.mburano, w.zheng}@westernsydney.edu.au, qiutie@tju.edu.cn

*Abstract*—Infrastructure networks such as the Internet backbone and power grids are essential for our everyday lives. With the prevalence of cyber-attacks on them, measuring their robustness has become an important issue. To date, many robustness metrics have been proposed. It is desirable for a robustness metric to possess the following three properties: considering global network topologies, strictly increasing upon link additions, and having a quadratic complexity in terms of the number of nodes on sparse networks. This paper proposes to use Average Network Flow (ANF) as a robustness metric, and proves that it increases strictly, and gives an algorithm to compute ANF with a quadratic complexity by leveraging Gomory-Hu trees. Thus, with ANF intrinsically considering global network topologies, ANF is unveiled to be a new robustness metric satisfying those three properties. Moreover, this paper compares ANF with seven existing representative metrics, showing that each metric has its own characteristics, so there is no silver bullet in measuring network robustness and it is recommended to apply several metrics together to gain a comprehensive view. Finally, by experimenting on the scenarios in which network topologies preserve the same numbers of nodes and links, some interesting behaviors of robustness metrics are reported.

## I. INTRODUCTION

OUR modern world is underpinned by infrastructure networks such as the Internet backbone, wireless mesh networks, power grids, etc. The robustness of these infrastructure networks is of great importance since they have become popular targets of cyber-attacks in recent years [1-4]. For example, the 'WannaCry' cyber-attack on UK's National Health Service network in 2017 [5] and the ransomware cyber-attack on Colonial Pipeline's oil pipelines in 2021 [6] were well-known events. The popularity of cyber-attacks on infrastructure networks is due to the following reasons: (1) security was typically not a major concern in the deployment stage, so the hardware and software in them may be insecure; (2) the impact of attacking infrastructure networks is high; (3) patching the hardware and software in infrastructure networks is difficult and slow due to the large scale and complexity of these networks.

Network robustness means the capability of a network to withstand failures or attacks [7]. Suitable robustness metrics are needed to study network robustness. Many robustness metrics have been proposed so far. The most basic metric is the average node degree [8] since robustness highly depends on the number of links present on a network. However, the average degree is a very rough metric, as it gives little consideration to the network topology (i.e., how links are arranged to connect nodes). Overcoming the roughness of average node degree, the metrics of Node Connectivity and Edge Connectivity [8] appeared to consider the network topology. However, their disadvantage is that local network topology is considered, resulting in that the Node/Edge Connectivity of a small portion of the network can dominate the entire network's Node/Edge Connectivity.

To further improve the granularity of robustness measures, metrics considering global network topologies have been proposed. Many such metrics are derived from graph spectrum [9], which is the set of eigenvalues of a graph's adjacency matrix or Laplacian matrix. It was an interesting discovery that the graph spectra from both adjacency matrix and Laplacian matrix are closely related to network robustness. Notable examples of such metrics include Algebraic Connectivity [10], Spectral Radius [11], the Number of Spanning Trees [12], Natural Connectivity [13], and Effective Graph Resistance [14]. All these five metrics perform well in terms of being sensitive to network topology changes. That is, different graphs with the same number of nodes and the same number of edges will mostly receive a different robustness value under these five metrics. Furthermore, the last three of these five metrics satisfy the 'strictly increase' property [15], which means that whenever a new edge is added to a graph, the value of the metric will increase strictly. Note that the first two of these five metrics have been shown not to satisfy this property. This is mainly because the first two only involve one eigenvalue from the graph spectrum in the metric calculation, while the last three involve all eigenvalues from the graph spectrum. The sensitivity of these graph-spectrum-based metrics comes at a price: the complexity of calculating them on general graphs is high, standing at $O(n^3)$, where $n$ is the number of nodes in a graph. If the graph is sparse, that is, $n$ is in a linear relationship with the number of edges (which is true for most infrastructure networks), then the complexity can be reduced to $O(n^2)$ [16].

A widely-accepted robustness metric that is not based on graph spectrum but considers the global network topology is the so-called $R$ [17]. It was inspired by the Percolation Theory [7], in which the size of the Largest connected Component (LC) in a network is a key indicator of how well a network is connected. The metric $R$ is obtained by considering the scenario that a network is attacked iteratively, and in each round, the node with the highest degree and its incident edges are removed from the network until all nodes are removed. The $R$ value is then calculated by averaging the ratio of the LC size and the network size after each round of node removal. Since $R$ considers the LC size in each round of node removal, its value is sensitive to network topology change, which is an advantage for a robustness metric. However, $R$ fails to increase strictly whenever an edge is added to a graph, which is a disadvantage for a robustness metric. As to be analyzed in Section III, calculating $R$ has the complexity $O(nm)$, where $n$ is the number of nodes, and $m$ is the number of edges in a graph. This complexity becomes $O(n^2)$ when the graph is sparse, so it is on a par with those graph-spectrum-based metrics in terms of complexity.

This paper aims to propose a new metric that is not based on graph spectrum but has the following three desirable properties discussed previously: (1) considering the global network topology; (2) increasing strictly when a new edge is added; (3) the computational complexity is $O(n^2)$ for sparse graphs. Our answer to this aim is to use the *Average Network Flow* (ANF) as a robustness metric, where ANF is defined as the average of the maximum network flows among all pairs of nodes in a network. Firstly, since ANF depends on the network flows among all node pairs, it considers the global network topology. Secondly, based on its definition, we prove that ANF increases strictly upon edge additions. Thirdly, we propose an algorithm for calculating ANF that achieves the complexity of $O(n^2)$ when the network is sparse and unweighted (i.e., each edge in the network has a unit capacity). The basic idea of this algorithm is to compute the Gomory-Hu tree [18] of an unweighted network first and then calculate the ANF based on this Gomory-Hu tree. The key for achieving the $O(n^2)$ complexity is to leverage a very recent algorithm in 2020 [19] for computing Gomory-Hu trees, which has the complexity of $O(n^{1.5})$. We will detail the concept of Gomory-Hu tree and its computation algorithms in Section II Preliminaries.

Note that, although ANF is quite a natural concept, it has not appeared in the literature previously as far as we know. However, we are aware of a similar concept called 'average connectivity of a graph' [20], which is defined as the *average node connectivity* of all node pairs in a graph. It is easy to see that the unweighted version of ANF is equivalent to the *average edge connectivity* of a graph, so it is different from the 'average connectivity' proposed in [20]. More importantly, we want to emphasize that the main contribution of this paper is not to propose the ANF concept, but the $O(n^2)$ algorithm for computing ANF by leveraging Gomory-Hu trees. Without this algorithm, a straightforward computation of ANF on unweighted and sparse networks will still take $O(n^3)$ time.

Moreover, this paper conducts experiments to compare the following seven existing robustness metrics as well as ANF: Node Connectivity (NodeC), Critical Fraction (CF) [7], Algebraic Connectivity (AC), the Number of Spanning Trees (NST), Natural Connectivity (NatC), Effective Graph Resistance (EGR), and $R$. The choice of these seven metrics is due to their wide adoption in measuring network robustness and their possession of clear intuition. We briefly discuss these seven metrics in Section III of this paper. Our comparisons on these robustness metrics shed light on three aspects that have not been adequately examined in the existing comparison works, such as [15, 21-23]. These three aspects are briefed below.

- The impact of four important graph properties (Degree Variance, Average Shortest Path Length, Average Clustering Coefficient, and Assortativity Coefficient) [24] on robustness metrics. We show which graph property affects a robustness metric the most.
- The tendency exhibited by a metric towards measuring the robustness under random failures or under targeted attacks. Here 'random failures' means that nodes in a network have an equal probability to fail, and 'targeted attacks' means that important nodes (e.g., nodes with large degrees) are removed first. We show which metrics tend to measure network robustness under random failures and under targeted attacks respectively.
- The performances of three popular graph models under those aforementioned eight metrics. These three models include Erdos-Renyi (ER) for random networks [25], Barabasi-Albert (BA) for scale-free networks [26], and Watts-Strogatz (WS) for small-world networks [27]. We generate a large number of graphs under each graph model, and then calculate, on average, which graph model performs better under a robustness metric.

The rest of this paper is organized as follows. Section II introduces the preliminaries needed by this paper. Section III discusses related work and gives the formula for calculating the seven existing robustness metrics experimented in this paper. Section IV details our algorithm for calculating ANF and analyzes its complexity. Section V describes our experimental setup and methodology, and Section VI presents our experimental results. Finally, Section VII concludes this paper.

## II. PRELIMINARIES

This section introduces the basic concepts needed by this paper, including graph spectrum, the four important graph properties mentioned in Section I, network flows, and Gomory-Hu tree.

### A. Graph and Graph Spectrum

In this paper, a network is modeled by a simple undirected graph $G=(V, E)$, where $V$ is the set of vertices (or nodes) labeled from 1 to $n$, and $E$ is the set of links (or edges). We let $n = |V|$, and $m = |E|$.

The *adjacency matrix* of a graph $G$, denoted by $A(G)$, is a zero-one $n \times n$ square matrix where its element $A_{ij}$ equals one if an edge exists between vertex $i$ and vertex $j$, and equals zero otherwise.

The *Laplacian matrix* of a graph $G$, denoted by $L(G)$, is the difference between the *degree matrix* of $G$, denoted by $D(G)$,

and $A(G)$. That is, $L(G) = D(G) – A(G)$. Here $D(G)$ is the diagonal matrix containing the degree of each vertex in its diagonal.

The *spectrum* of a graph $G$ is the set of eigenvalues of $A(G)$ or $L(G)$. In this paper, the spectrum of $A(G)$ is denoted by $\{\mu_1,…, \mu_n\}$, and the spectrum of $L(G)$ is denoted by $\{\lambda_1, …, \lambda_n\}$. Without loss of generality, we assume $\mu_1 \le \mu_2 … \le \mu_n$, and $\lambda_1 \le \lambda_2 … \le \lambda_n$.

### B. Four Important Graph Properties

*Degree Variance:* is the variance of all node degrees in a graph. This property is considered a measure of a graph heterogeneity.

*Average Shortest Path Length:* is the average of the shortest path lengths of all nodes pairs in a graph [24]. This property is used to measure the efficiency of information transfer on a network.

*Average Clustering Coefficient*: is the average of the local clustering coefficients of all nodes in a graph [24]. This property reflects the clustering level of the entire graph.

*Assortativity Coefficient*: is the Pearson correlation coefficient of degrees between pairs of linked nodes in a graph [24]. This property reflects the general tendency of every pair of linked nodes to have similar node degrees in a graph.

### C. Network Flows

A *flow network* is a graph $G=(V, E)$ where each edge $(u, v) \in E$ has a capacity $c(u, v) \ge 0$, and there are a source node $s$ and a sink node $t$ in $V$. A *flow* in a network $G$ is a real-valued function $f: V \times V \to \mathbb{R}$ that satisfies the following two constraints: (1) for any $u, v \in V$, $0 \le f(u, v) \le c(u, v)$, which is called the Capacity Constraint; and (2) for any $u \in V – \{s, t\}$, $\sum_{v \in V} f(v, u) = \sum_{v \in V} f(u, v)$, which is called the Flow Conservation Constraint. Note that this paper will use 'capacity' and 'weight' interchangeably due to the legacy in the literature.

The value of a feasible flow from $s$ to $t$ on $G$, denoted by $|f|$, is defined as: $|f| = \sum_{v \in V} f(s, v)$, which means the total flow out of source node $s$. The maximum flow problem is to find the maximum $|f|$. For convenience, this paper denotes the maximum $|f|$ between a source $s$ and a sink $t$ in a network $G$ by $F(G, s, t)$.

Many algorithms have been proposed to calculate $F(G, s, t)$ so far. According to the Wikipedia entry [28], the most efficient one is Orlin's algorithm [29], which has the complexity of $O(nm)$. If we use Orlin's algorithm to calculate ANF naively, since there are altogether $n(n-1)/2$ pairs of source and sink in a network, the overall complexity becomes $O(n^3m)$, which is too high. To overcome this problem, this paper resorts to Gomory-Hu tree [18] to bring down the complexity.

### D. Gomory-Hu Tree

Given an undirected graph $G=(V, E)$ with edges having non-negative capacities, a Gomory-Hu tree [18] of $G$, denoted by $T$, is a weighted tree also taking $V$ as the node set and satisfying the following two properties: (1) for any source/sink pair $(s, t) \in V$, if we remove the least weighted edge on the path from $s$ to $t$ in $T$, the $s$-$t$ cut [30] obtained on the $V$ of $T$ is a minimum $s$-$t$ cut on the $V$ of $G$; (2) the weight of this least weighted edge equals $F(G, s, t)$ on $G$. Note that it has been proved that a Gomory-Hu tree does not always exist on directed graphs [31], so it is defined on undirected graphs only. From the properties of the Gomory-Hu tree, we see that the maximum flows of all node pairs on a graph can be easily obtained if the Gomory-Hu tree on this graph is available.

An interesting result from [18] is that a Gomory-Hu tree can be computed with $n$-1 maximum flow computations, not the naïve $n(n$-$1)/2$ computations. This means that a Gomory-Hu tree can be computed with time complexity $O(n^2m)$ if Orlin's algorithm is used. Especially for unweighted graphs, a very recent result [19] shows that a Gomory-Hu tree can be computed in $O(m^{3/2+o(1)})$ time. When the graph is sparse as well, the complexity becomes $O(n^{3/2+o(1)})$, which greatly reduces the complexity of obtaining the maximum flows among all node pairs. Therefore, the basic idea of this paper is to use the algorithm from [19] to calculate the Gomory-Hu tree of an unweighted graph, and then calculate the ANF based on this Gomory-Hu tree.

### III. RELATED WORK

This section discusses the seven representative robustness metrics compared in this paper, and also surveys other comparison works existing in the literature.

### A. Seven Existing Robustness Metrics

We first note that all seven metrics except EGR below have a positive relationship with network robustness (i.e., the higher the value, the more robust a network is). Thus, for EGR, we calculate their reciprocals instead in our experiments so as to interpret their results consistently with the other six metrics.

*Node Connectivity* (NodeC) [8] is the minimum number of nodes whose removal disconnects a graph. Note that Edge Connectivity is the minimum number of edges whose removal disconnects a graph. Since Edge Connectivity is close to and highly correlated with Node Connectivity, this paper does not include Edge Connectivity in the comparison.

*Critical Fraction* (CF) [7] is a robustness metric from Percolation Theory, in which the size of the Largest connected Component (LC) of a network is a key indicator of how well a network is connected [7]. Specifically, CF is defined as the fraction of nodes whose random removal makes the LC almost disappear (i.e., the ratio of the LC size and the entire network size is almost zero). It is discovered that, regardless of the node degree distribution in a network, the CF of a network can be calculated using the formula below, where $\langle k \rangle$ denotes the average node degree and $\langle k^2 \rangle$ denotes the average of the degree square of all nodes.

$$CF = 1 - \frac{1}{\frac{\langle k^2 \rangle}{\langle k \rangle} - 1} \qquad (1)$$

Note that this formula holds when $n$ approaches infinity, but when $n$ is finite, this formula is still true approximately. Since the above formula is obtained by assuming random node removals, CF is a metric for measuring network robustness under random failures.

*Algebraic Connectivity* (AC) [10] is the second smallest eigenvalue of the Laplacian matrix of a graph. That is, it equals $\lambda_2$. Note that $\lambda_1$ is always zero since the Laplacian matrix of a graph is symmetric and positive semidefinite. AC is considered a robustness metric because it has the following two properties [15]: (1) if $\lambda_2=0$, the graph is disconnected; if $\lambda_2>0$, the graph is connected; (2) $\lambda_2$ provides a lower bound for NodeC and edge connectivity, i.e., $\lambda_2 \leq$ NodeC $\leq$ edge connectivity.

*The Number of Spanning Trees* (NST) [12] is the total number of spanning trees in a graph. According to Kirchhoff's matrix-tree theorem [9], for a connected graph $G$, NST is equal to the normalized product of all positive eigenvalues of $L(G)$. That is,

$$NST = \frac{1}{n}\prod_{i=2}^{n} \lambda_i \quad (2)$$

In a network with probabilistic link losses, the probability that there exists a path between any pair of nodes is equal to the probability of the existence of a spanning tree. Thus, NST is used as a robustness metric in many works.

*Natural Connectivity* (NatC) [13] is the normalized sum of the scaled numbers of closed walks of all lengths in a graph $G$. It happens that NatC can be calculated using the eigenvalues of $A(G)$ as follows:

$$NatC = \ln\left[\frac{1}{n}\sum_{i=1}^{n} e^{\mu_i}\right] \quad (3)$$

Since the number of closed walks is a key reflection on the number of alternative paths in a graph, NatC has been used in several works to measure network robustness.

*Effective Graph Resistance* (EGR) [14] is the sum of the effective resistances of all pairs of nodes in a graph $G$, which is deemed as an electrical circuit with all edges having a unit resistance. It is proven that EGR can be calculated using the eigenvalues of $L(G)$ as follows:

$$EGR = n\sum_{i=2}^{n} \frac{1}{\lambda_i} \quad (4)$$

Intuitively, the higher the EGR is, the circuit is more in the form of series rather than parallel, so the network is less robust. Thus, EGR has an inverse relationship with network robustness.

*R* [17], similar to CF, is a robustness metric inspired by Percolation Theory, but different from CF, $R$ is applied in the scenario of targeted attacks. Recall that CF only reflects network robustness based on the case of a disappeared LC. That is, it ignores the cases in which a network still has some connectivity before the LC disappears. To have a fine-grained assessment of network robustness, $R$ considers a network being attacked iteratively, and in each round, the node with the highest degree and its incident edges are removed from the network until all nodes are removed. Based on this scenario, $R$ is calculated by averaging the ratio of the LC size and the network size after each round of node removal. Specifically, given a network $G$, the formula of calculating $R$ is given below:

$$R = \frac{1}{n}\sum_{i=1}^{n} \frac{LC(G_i)}{n} \quad (5)$$

Here, $G_i$ denotes the network after the $i$-th round of node removal, and $LC(G_i)$ denotes the LC size in $G_i$, and the $1/n$ at the beginning averages the results of all rounds. Due to its fine granularity, $R$ has been used in several works such as [32, 33] to measure network robustness.

Since the time complexity of calculating $R$ has not been given in the literature, we analyze it as follows. The computation of $R$ involves the following main steps in each round of node removal:
1. Computing the connected components, which takes $O(n+m)$ time [30].
2. Finding the LC, which takes $O(n)$ time.
3. Finding the highest degree node in the network, which takes $O(n)$ time.

Thus, by combining the total $n$ rounds of node removals, the entire complexity of computing $R$ is:
$O(n(n+m+n+n)) = O(3n^2 + nm) = O(nm)$.

### B. Other Comparison Works

While many metrics on network robustness have been proposed so far, several comparison works [15, 21-23] on robustness metrics have also appeared. Here we briefly survey these works and also point out our paper's differences from these works. A general difference of our work is that our comparison focuses on how the robustness metrics behave on different graphs with the same $n$ and $m$, while the existing works are mainly concerned with the behavior of the metrics when $m$ changes.

In [15], a survey on thirteen network robustness metrics based on classical graph properties and graph spectra is given. This survey presents detailed discussions and comparisons on robustness metrics under the following three criteria: the clarity of intuition, the 'strictly increase' property, and the number of redundant paths between a pair of nodes. The experiments are only conducted on some baseline graphs, but not on graphs under complex network models such as those used in our paper.

In [21], the focus is on evaluating the metrics on their accuracy in measuring network robustness against targeted attacks. More than ten robustness metrics are compared, and experiments are conducted on ten baseline graphs as well as synthetic graphs under Gilbert graphs model, Waxman graphs model, and Gabriel graphs model, which are different from the models used in this paper. The results shed light on which robustness metric is more accurate on a certain graph model.

In [22], the comparisons focus on the following two aspects of robustness metrics: the sensitivity in measurement and the guidance in optimizing network robustness. Regarding the sensitivity, edges are added or deleted from networks to observe the changes of the robustness metrics. Regarding the guidance, networks are optimized with a metric as the objective, and the optimization process uses the Hill Climbing technique [34]; then, the optimized networks are evaluated to see how robust they are. The experiments are only conducted on the scale-free networks (under the Barabasi-Albert model) since the paper deems that scale-free networks simulate real-world networks very well.

In [23], the study of robustness metrics is conducted on fifteen real-world telecommunication networks. The main results include the following. First, the paper shows the relationships between robustness metrics and well-known graph properties such as Average Shortest Path Length and Assortativity Coefficient. Note that, in [23], the relationships are discussed simply by observing the curves of metrics and properties in the data plots, while our paper shows relationships by calculating the Spearman's Correlation Coefficient. Second, the paper shows which robustness metric is relatively high on a certain real-world network. Note that our paper also conducted a similar study, but we showed which robustness metric is relatively high on a certain graph model.

## IV. Algorithm For Average Network Flow

In this section, we first give the formula for calculating ANF and prove that ANF satisfies the 'strictly increase' property, and then detail our algorithm, and finally prove the time complexity of our algorithm is $O(n^2)$ on sparse networks.

### A. ANF and 'Strictly Increase' Property

Given an undirected flow network $G=(V, E)$ where each edge has a non-negative capacity, the ANF on $G$, denoted by ANF($G$), is the average of the maximum network flows among all pairs of distinct nodes in $G$. Given a source node $s$ and a sink node $t$, the maximum flow between $s$ and $t$ is denoted by $F(G, s, t)$ as mentioned in Subsection II.C. Since we assume $G$ is undirected, we can ignore the order of $s$ and $t$ here. Thus, there are altogether $\frac{n(n-1)}{2}$ node pairs in $G$, and then the formula for calculating ANF is given as follows:

$$ANF(G) = \frac{2}{n(n-1)} \sum_{(s,t)\in V\times V,\ s<t} F(G,s,t) \qquad (6)$$

Note that here we use '$s < t$' to indicate that we only count a node pair once by ignoring the order of nodes, utilizing our assumption that nodes are labelled from 1 to $n$ in this paper.

Now we prove that ANF possesses the 'strictly increase' property, which is stated in the following theorem.

**Theorem 1.** *Suppose an undirected flow network $G=(V, E)$ is incomplete, i.e., has some missing edges. If we add any missing edge, say $(u, v)$, with a positive capacity $c(u,v)>0$, to $G$, we obtain a new flow network $G'$ $(V, E \cup \{(u,v)\})$. Then, we have ANF($G'$)>ANF($G$).*

**Proof:** Since ANF is the average of maximum flows of all node pairs, we consider every node pair in ANF($G'$) and ANF($G$) in the proof below. In considering the maximum flow between a node pair, we resort to the *Max-Flow Min-Cut Theorem* [30], which states that the maximum flow between a node pair equals the capacity of the minimum cut that separates that node pair. Hereafter, we denote the minimum cut for a node pair $(s, t)$ in a graph $G$ by MinC($G, s, t$), and the capacity of MinC($G, s, t$) by $c$(MinC($G, s, t$)). Also, refer to Fig. 1 for an exemplar flow network where a new edge $(u, v)$ is added.

First, for a node pair $(s, t)$ that is different from $(u, v)$, we show $F(G', s, t) \geq F(G, s, t)$. Suppose this is not true, which means that $c$(MinC($G', s, t$)) < $c$(MinC($G, s, t$)). There can be the following two cases.

- *MinC($G', s, t$) includes the edge $(u, v)$*: Since MinC($G', s, t$) separates $s$ and $t$ in $G'$, MinC($G', s, t$) – $\{(u, v)\}$ separates $s$ and $t$ in $G$ as well, implying that MinC($G', s, t$) – $\{(u, v)\}$ is a cut for $s$ and $t$ in $G$. However, we have $c$(MinC($G', s, t$) – $\{(u, v)\}$) < $c$(MinC($G', s, t$)) < $c$(MinC($G, s, t$)). This contradicts that MinC($G, s, t$) is a minimum cut.
- *MinC($G', s, t$) does not include the edge $(u, v)$*: Since MinC($G', s, t$) separates $s$ and $t$ in $G'$, it separates $s$ and $t$ in $G$ as well, implying that it is a cut for $s$ and $t$ in $G$. However, we have $c$(MinC($G', s, t$)) < $c$(MinC($G', s, t$)). This contradicts that MinC($G, s, t$) is a minimum cut.

Since both cases above lead to contradictions, we have proved that $F(G', s, t) \geq F(G, s, t)$.

Second, for the node pair $(u, v)$ itself, we show that $F(G', u, v) = F(G, u, v) + c(u, v) > F(G, u, v)$. Since MinC($G', u, v$) must include the edge $(u, v)$ in it, any MinC($G', u, v$) – $\{(u, v)\}$ is a minimum cut for $u$ and $v$ in $G$, and any MinC($G, u, v$) ∪ $\{(u, v)\}$ is a minimum cut for $u$ and $v$ in $G'$. Thus, we have $F(G', u, v) = F(G, u, v) + c(u, v) > F(G, u, v)$.

Now we have shown that for the node pair $(u, v)$, $F(G', u, v) > F(G, u, v)$, and that for a node pair $(s, t)$ other than $(u, v)$, $F(G', s, t) \geq F(G, s, t)$. Therefore, combining all node pairs together, we have ANF($G'$) > ANF($G$). ∎

### B. Algorithm Description

If we compute the ANF on a graph $G$ using the Formula (6) given in the previous subsection, it will involve $n(n-1)/2$ computations of $F(G, s, t)$, which is not efficient. Fortunately, as mentioned in Subsection II.D of this paper, the Gomory-Hu tree of $G$, denoted by $T$, has the property that for any source/sink pair $(s, t)$ in $G$, the weight of the least weighted edge on the path from $s$ to $t$ in $T$ equals $F(G, s, t)$. Thus, if we have $T$ available, we can compute ANF($G$) by finding out those least weighted edges on the path between all node pairs in $T$, and using the weights of those edges as $F(G, s, t)$.

As shown in a very recent paper [19], $T$ can be computed in an efficient $O(m^{3/2+o(1)})$ time if each edge in $G$ has a unit capacity. To leverage this result in calculating ANF, this paper assumes unit capacity on $G$'s edges.

Given $T$, if we find out the least weighted edge on the path between each node pair one by one, it requires iterating through all $n(n-1)/2$ node pairs, and for each node pair, an $O(n)$ complexity to determine the least weighted edge. This will result in an $O(n^3)$ complexity, which is not efficient as well. To overcome this problem, this paper presents a Depth First Search (DFS) based algorithm that can find out the least weighted edges from a single source node to all other nodes in $O(n)$ time.

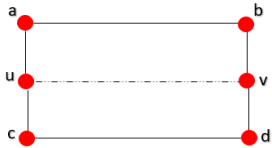

Fig. 1: An exemplar flow network

**Algorithm 1:** Overall algorithm for computing ANF

**Input:** $G=(V, E)$
**Output:** ANF
1. Set the capacity of every edge in $E$ to one;
2. Compute $T(G)$ by calling the algorithm from [19];
3. Initialize $flow\_array[n, n]$ to store maximum flows among all node pairs;
4. **for** $s = 1$ to $n$ **do**
    Call the SSLT Algorithm to compute $flow\_array[s, x]$, where $x$ ranges from 1 to $n$;
   **end for**
5. $total\_flow = 0$;
6. **for** $s = 1$ to $n - 1$ **do**
      **for** $t = s + 1$ to $n$ **do**
          $total\_flow = total\_flow + flow\_array[s, t]$;
      **end for**
   **end for**
7. ANF = $2 \times total\_flow / n / (n - 1)$

**Algorithm 2:** The SSLT algorithm

**Input:** $T$, $current$, $parent$
**Output:** Filling in $flow\_array[n, n]$ during recursive calls

**for** $nbr$ in $T[current]$ **do**
    **if** ($nbr ==$ $parent$)
        **continue**;
    **end if**
    **if** ($current == s$)
        $flow\_array[s][nbr] = T[s][nbr].weight$;
    **else if** ($T[current][nbr].weight < flow\_array[s][current]$)
        $flow\_array[s][nbr] = T[current][nbr].weight$;
    **else**
        $flow\_array[s][nbr] = flow\_array[s][current]$
    **end if**
    // traverse the next node by DFS
    SSLT($T$, $nbr$, $current$)
**end for**

We call this algorithm the *Single Source Least Weight* (SSLT) algorithm by following the naming convention of the *Single Source Shortest Path* algorithm [30]. With this SSLT algorithm, we can use each node in $T$ as the source node and obtain its least weighted edges to all other nodes in $O(n)$ time. Thus, we can compute the least weighted edges among all node pairs in $O(n^2)$ time on a Gomory-Hu tree.

With the above said, we give the pseudocode of our overall algorithm for computing ANF in Algorithm 1, which uses the SSLT algorithm as a subroutine. The pseudocode of SSLT is given in Algorithm 2. In Algorithm 1, we first call the algorithm from [19] to compute $T$. Then, using each node in $V$ as the source node, the SSLT algorithm is called to compute the least edge weight between this source node and the other nodes. The results are saved in an $n \times n$ array named $flow\_array$. Finally, the maximum flows among all node pairs are summed up and the ANF is calculated.

Algorithm 2 presents the pseudocode of a recursive function call SSLT($T$, $current$, $parent$), which traverses the Gomory-Hu tree by DFS and meanwhile computes the least edge weights between a single source node and the other nodes. For the function parameters, $T$ is the known Gomory-Hu tree of a graph with the given source node as the root node, $current$ is the current node that the recursive function is visiting, and $parent$ is the parent node of the current node in $T$. Note that, in the step 4 of Algorithm 1 (the loop using every node as the source node $s$), the SSLT should be called as SSLT($T$, $s$, -1), where '-1' is used to indicate that $s$ does not have a parent node, i.e., it serves as the root node of $T$ in this iteration.

Inside the function body of Algorithm 2, we use $T[u]$ to denote the set of node $u$'s neighbor nodes in $T$, and use $nbr$ to denote a neighbor node in $T[u]$, and use $T[u][v]$ to denote the edge $(u, v)$ in $T$, and use $T[u][v].weight$ to denote the edge weight of $T[u][v]$. The function SSLT($T$, $current$, $parent$) traverses $T$ recursively using DFS and meanwhile finds out the least edge weights between a source node and the other nodes in $T$. The basic idea is:

- At any current node during the tree traversal, the least edge weight is stored in $flow\_array[s][current]$.

- For the next node (i.e., $nbr$) to determine the least edge weight, if $T[current][nbr].weight$ is less than $flow\_array[s][current]$, then $flow\_array[s][nbr]$ is set to $T[current][nbr].weight$, otherwise $flow\_array[s][nbr]$ is set to $flow\_array[s][current]$.

### C. Complexity Analysis

This subsection analyzes the time complexity of our algorithm for computing ANF. The result is presented in the following theorem.

**Theorem 2.** *The time complexity of Algorithm 1 is $O(m^{3/2+o(1)} + n^2)$, which is $O(n^2)$ when a network is sparse.*

**Proof:** In Algorithm 1, the main computations are done in Steps 3, 4, and 6. Step 3 computes the Gomory-Hu tree by calling the algorithm from [19], which has the complexity $O(m^{3/2+o(1)})$. Step 4 calls our SSLT algorithm $n$ times. The SSLT algorithm exactly traverses each edge in a Gomory-Hu tree once by following DFS. Since $m = n - 1$ in a tree, SSLT has the complexity of $O(n)$. Thus, the entire Step 4 has the complexity of $O(n^2)$. Step 6 uses two 'for' loops to sum up the maximum flows among all distinct node pairs, so it has the complexity of $O(n^2)$.

Combining the computations in Steps 3, 4, and 6, the complexity of Algorithm 1 is $O(m^{3/2+o(1)} + n^2 + n^2) = O(m^{3/2+o(1)} + n^2)$.

We can further discuss this complexity in the following two cases. In case 1, when the network is close to a complete graph, $m$ is in the order of $n^2$. Thus, the term $m^{3/2+o(1)}$ dominates the complexity, and the total complexity becomes $O(n^{3+o(1)})$. In case 2, when the network is sparse, the total complexity becomes $O(n^{3/2+o(1)} + n^2) = O(n^2)$. ∎

## V. EXPERIMENTAL SETUP AND METHODOLOGY

This section describes our experimental settings and methodology. Especially, we introduce an algorithm for generating a set of graphs with different degree distributions to experiment the general behavior of robustness metrics.

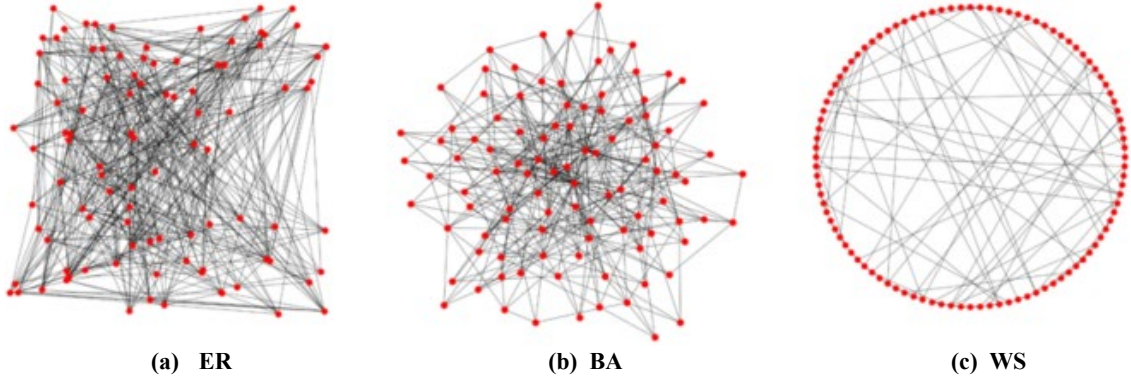

(a) ER  (b) BA  (c) WS

Fig. 2: Exemplar graphs of ER, BA, and WS models

## A. Spearman's Rank Correlation Coefficient

In this paper, Spearman's Rank Correlation Coefficient [35] is adopted to show the relationship between two robustness metrics or between a robustness metric and a graph property. Here a rank correlation coefficient instead of a linear correlation coefficient is chosen because we want to focus on how robustness metrics rank various graphs, and also different robustness metrics are not necessarily linearly correlated. The Spearman's Rank Coefficient ranges from -1 to 1. If its value is larger than zero, it shows a positive correlation; if smaller than zero, a negative correlation; and if zero, no correlation.

We typically generate a set of graphs, and then calculate robustness metrics and graph properties on this set of graphs, and finally obtain the Spearman's Coefficient between any two measures.

## B. Generating Graphs under Three Well-known Graph Models

In our experiments, we use the graphs from the following three well-known models for large-scale networks: Erdos-Renyi (ER) for random networks [25], Barabasi-Albert (BA) for scale-free networks [26], and Watts-Strogatz (WS) for small world networks [27]. To provide intuition, exemplar graphs from these three models are illustrated in Fig. 2. We generate the graphs under these three models by calling APIs from the NetworkX [36] library, which is a widely-used Python package for analyzing large-scale networks today. Under each graph model, we generated 50 graphs randomly. Then, robustness metrics and graph properties are calculated on this set of 50 graphs. With these values obtained, the Spearman's Correlation Coefficients between metrics or properties are calculated.

To make our comparisons fair, the graphs under all three models are generated with the same $n$ and the same $m$. Specifically, we use $n=200$ and $m=800$ in this paper. The use of the same $n$ and $m$ is especially important because we have a set of experiments about which graph model performs better on a robustness metric. We note that, even under the same $n$ and $m$, a huge number of non-isomorphic graphs still exist. With $n=5$ and $m=5$ as an example, some graphs satisfying this condition are illustrated in Fig. 3. We see that network topologies can still vary significantly. Thus, it is meaningful to experiment on how robustness metrics behave when $n$ and $m$ remain the same.

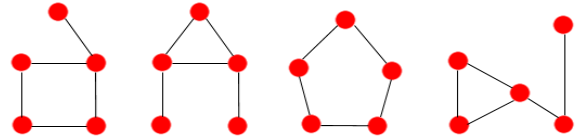

Fig. 3: Four non-isomorphic connected graphs with $n=5$ and $m=5$

## C. Generating A Set of Graphs with Different Degree Distributions

For the three graph models above, all graphs under the same model have a very similar node degree distribution. It is known that the degree distribution in ER model is Poisson when the number of nodes approaches infinity; the degree distribution in BA model follows power law; and the degree distribution in WS model can be a little broad, but is heavy-tailed in general. Since the degree distributions in these three graph models are quite fixed, it is helpful to introduce a set of graphs with different degree distributions, so that we can learn the general behaviour of robustness metrics.

---

**Algorithm 3:** Generating a set of graphs with different degree distributions

**Input:**
  $n$: the number of nodes (should be even)
  $m$: the number of edges ($2m$ should be a multiple of $n$)
  $num\_graphs$: the number of graphs in the set

**Output:**
  $graphs[1:num\_graphs]$: a set of $num\_graphs$ graphs

1. Calculate the average node degree: $d = 2m/n$;
2. **for** $i = 1$ to $n$ **do**
     $degrees[i] = d$;
   **end for**
3. $graphs[1]$ = a random graph with $degrees$ as degree sequence;
4. Calculate the total number of possible degrees to move from left nodes to right nodes when generating a graph:
   $deg\_delta = (m - n/2) / num\_graphs$;
5. **for** $i = 2$ to $num\_graphs$ **do**
     **for** $j = 1$ to $deg\_delta$ **do**
       Randomly pick a node $l$ from left nodes;
       $degrees[l] = degrees[l] - 1$;
       Randomly pick a node $r$ from right nodes;
       $degrees[r] = degrees[r] + 1$;
     **end for**
     $graphs[i]$ = a random graph with $degrees$ as degree sequence;
   **end for**

For this purpose, we designed an algorithm for generating a set of graphs with different degree distributions and also preserving the same $n$ and $m$. The pseudocode of this algorithm is presented in Algorithm 3. This algorithm is explained as follows. Step 1 calculates the average node degree ($d$) based on $n$ (needs to be even) and $m$. Step 2 initializes a degree sequence (*degrees*[1:$n$]) with every degree being $d$. Step 3 generates the first graph in this graph set, which is a random $d$-regular graph based on the degree sequence stored in *degrees*[1:$n$]. Note that, although this is a $d$-regular graph, there are still many possible instances. We just randomly generate one here. This can be easily done by resorting to NetworkX, which includes a method for generating a graph randomly based on a degree sequence. This method will be denoted by *rand_seq_graph*() hereafter.

In Step 4, the nodes are divided into two groups: the left group and the right group, with the left group containing the nodes from 1 to $n/2$, and the right group containing the nodes from $n/2+1$ to $n$. Then, the total number of possible degrees to move from left nodes to right nodes to change Degree Variances is calculated. In Step 5, a loop of *num_graphs* – 1 iterations is executed, where *num_graphs* denotes the total number of graphs to be generated in this graph set. The basic idea of this loop is to move degrees from the left nodes to the right nodes gradually to obtain degree sequences with larger and larger Degree Variances. When the Degree Variances are different, the degree distributions must be different. Thus, the graphs with different degree distributions are generated into this graph set. At the end of each iteration, the *rand_seq_graph*() is called with the current *degrees*[1:$n$] to generate a graph randomly. The number of degrees to move in each iteration is amortized, such that the last graph in this set will see the left nodes all having degree 1 and the right nodes having large degrees around $2d$, representing a graph with an extremely large Degree Variance.

## VI. EXPERIMENTAL RESULTS

This section presents our experimental results on the seven existing robustness metrics discussed in Section III and ANF, the metric proposed by us. As overviewed in Section I, our experiments cover three aspects on these eight metrics. Below, we detail the results for each aspect with one subsection.

### A. The Impact of Four Important Graph Properties on The Eight Robustness Metrics

Graphs have many properties defined on them. It is natural to ask how the increase/decrease of a graph property value will affect a robustness metric. In this subsection, we pick four important graph properties (Degree Variance, Average Shortest Path Length, Average Clustering Coefficient, and Assortativity Coefficient) that are relevant for infrastructure networks, and show their impacts on robustness metrics. Here the impact is measured by the Spearman's correlation coefficient between a graph property and a robustness metric. We present the results using four tables, with each table containing the results for one graph property. In each table, we present the results under four sets of graphs: DDD, BA, ER, and WS, where DDD denotes the set of graphs with different degree distributions, and BA denotes the set of graphs under Barabasi-Albert model, and ER denotes the set of graphs under Erdos-Renyi model, and WS denotes the set of graphs under Watts-Strogatz model.

TABLE I: CORRELATION RESULTS FOR DEGREE VARIANCE

|       | DDD     | BA      | ER      | WS      |
|-------|---------|---------|---------|---------|
| NodeC | -0.8267 | -0.3540 | -0.3747 | -0.3101 |
| CF    | 1.0000  | 1.0000  | 1.0000  | 1.0000  |
| AC    | -0.8291 | -0.0970 | -0.3562 | 0.4645  |
| NatC  | 0.9999  | 0.9546  | 0.9525  | -0.2981 |
| NST   | -1.0000 | -0.8581 | -0.8732 | -0.0269 |
| EGR   | -0.9998 | -0.6606 | -0.7321 | 0.3472  |
| R     | -0.9963 | -0.5613 | -0.4842 | -0.1134 |
| ANF   | -1.0000 | -0.7678 | -0.9747 | -0.9872 |

Table I gives the Spearman's coefficients between the *Degree Variance* (DV) and the eight robustness metrics under the four sets of graphs. The most obvious observation is that CF has a perfectly positive correlation with DV. This is because CF fully depends on $\langle k^2 \rangle$ and $\langle k \rangle$ according to its formula, and so does DV (which equals $\langle k^2 \rangle - \langle k \rangle^2$). As aforementioned, all graphs generated in our experiments have the same $n$ and $m$, so all graphs have the same $\langle k \rangle$. Thus, $\langle k^2 \rangle$ will fully decide the values of CF and DV according to their formulas, so they are perfectly correlated. Intuitively, this can also be explained. As discussed in the Related Work section, CF reflects the robustness under random failures. When DV is large (typically, a small portion of nodes have high degrees and a large portion of nodes have low degrees), failure nodes tend to be those nodes with low degrees, which explains the robustness of the network.

Other notable observations from Table I include:
- Very strong correlations of exactly or nearly ±1 are seen under the DDD graphs. This is because DDD graphs are specially generated with significant differences of DV among them. These values of exactly or nearly ±1 show that DV has a very strong impact on all the eight robustness metrics when $n$ and $m$ remain the same.
- Besides CF, NatC also has a positive correlation with DV in general, implying that the increase of DV tends to increase NatC. The other seven metrics all have a negative correlation with DV under all four sets of graphs, implying that the increase of DV tends to decrease these metrics.

TABLE II: CORRELATION RESULTS FOR AVERAGE SHORTEST PATH LENGTH

|       | DDD     | BA      | ER      | WS      |
|-------|---------|---------|---------|---------|
| NodeC | -0.7868 | 0.2340  | -0.2577 | -0.0010 |
| CF    | 0.9655  | -0.9489 | 0.3257  | -0.4652 |
| AC    | -0.8480 | -0.0423 | -0.3379 | -0.8315 |
| NatC  | 0.9659  | -0.8633 | 0.4646  | 0.9471  |
| NST   | -0.9655 | 0.8415  | -0.5003 | -0.8066 |
| EGR   | -0.9661 | 0.5956  | -0.4999 | -0.9564 |
| R     | -0.9594 | 0.5991  | -0.0815 | 0.0325  |
| ANF   | -0.9655 | 0.7510  | -0.3927 | 0.4335  |

Table II gives the Spearman's coefficients between the *Average Shortest Path Length* (ASPL) and the eight robustness metrics. From this table, we can mainly observe the following:
- ASPL also has a strong impact on the eight robustness metrics under DDD graphs, although its impact is a little weaker than DV.

TABLE III:
CORRELATION RESULTS FOR AVERAGE CLUSTERING COEFFICIENT

|       | DDD     | BA      | ER      | WS      |
|-------|---------|---------|---------|---------|
| NodeC | -0.7959 | -0.3065 | 0.0017  | -0.1187 |
| CF    | 0.8825  | 0.8578  | 0.3907  | -0.3862 |
| AC    | -0.783  | -0.0591 | 0.0563  | -0.7481 |
| NatC  | 0.8819  | 0.7928  | 0.3436  | 0.9034  |
| NST   | -0.8825 | -0.7914 | -0.2257 | -0.8646 |
| EGR   | -0.8838 | -0.6467 | -0.1346 | -0.9579 |
| R     | -0.878  | -0.632  | -0.1709 | 0.0051  |
| ANF   | -0.8825 | -0.7349 | -0.396  | 0.3605  |

- For most robustness metrics, the correlations are not consistent, with a positive correlation on some graph sets and a negative correlation on some other graph sets. This shows that there are no definite correlations between ASPL and these robustness metrics, and some other graph properties may play a stronger role in affecting the robustness metrics.

Table III gives the Spearman's coefficients between the *Average Clustering Coefficient* (ACC) and the eight robustness metrics. From this table, we can mainly observe the following:
- ACC also shows a strong impact on the eight robustness metrics under DDD graphs, although its impact is further weaker than ASPL.
- ACC generally exhibits a positive correlation with CF and NatC, and a negative correlation with the other seven metrics.

TABLE IV:
CORRELATION RESULTS FOR ASSORTATIVITY COEFFICIENT

|       | DDD     | BA      | ER      | WS      |
|-------|---------|---------|---------|---------|
| NodeC | 0.3321  | 0.0938  | -0.3161 | 0.1559  |
| CF    | -0.539  | -0.584  | 0.1292  | 0.0294  |
| AC    | 0.3497  | 0.0107  | -0.3294 | -0.1245 |
| NatC  | -0.5359 | -0.3887 | 0.3605  | 0.2258  |
| NST   | 0.539   | 0.5867  | -0.2829 | -0.2236 |
| EGR   | 0.5387  | 0.3947  | -0.3077 | -0.1661 |
| R     | 0.558   | 0.6179  | 0.2011  | 0.1486  |
| ANF   | 0.539   | 0.5489  | -0.1638 | -0.0646 |

Table IV gives the Spearman's coefficients between the *Assortativity Coefficient* (AsCo) and the eight robustness metrics. From this table, we can mainly observe the following:
- Unlike the previous three graph properties, AsCo does not show a strong impact on the eight metrics under the DDD graphs, implying that AsCo is not strongly related to network robustness in general.
- AsCo exhibits a consistent correlation with the robustness metric R under all four sets of graphs. This conforms with the conclusion drawn in an existing work [37] that the onion-like structure where nodes tend to connect to nodes with similar degrees contributes to a larger R.

*In summary, we can draw the following conclusions from the four tables in this subsection:*
- The impacts of the four graph properties on network robustness decrease in the following order: DV, ASPL, ACC, and AsCo, with DV being the strongest impact factor on these robustness metrics.
- The ANF metric proposed in this paper is consistently and negatively correlated with DV under all four sets of graphs, implying that ANF tends to be large in networks where nodes have similar degrees. On the other hand, ANF is not consistently correlated with the other three graph properties.
- In many cases, the correlations between the four graph properties and the eight robustness metrics are not consistent among different graph sets, implying that there are other properties also affecting the values of robustness metrics.

*B. The Tendency of Measuring Robustness under Random Failures or Targeted Attacks*

Network robustness has two main scenarios: random failures and targeted attacks. It has been shown that a network being robust under random failures may not be robust under targeted attacks, and vice versa [22]. Two robustness metrics of the eight experimented in this paper assume a scenario for being used: CF assumes random failures, and R assumes targeted attacks. However, the other six metrics do not need to assume a scenario. That is, they can be applied to both scenarios. Thus, it is interesting to see the tendencies exhibited by these seven metrics toward measuring robustness under random failures, as well as under targeted attacks.

In this subsection, we use CF as the standard metric for robustness under random failures, and use R as the standard metric for robustness under targeted attacks. We measure the tendencies of a metric toward random failures and targeted attacks by obtaining this metric's Spearman's correlation coefficient with CF and R, respectively. The measurements are conducted under four sets of graphs: DDD, BA, ER, and WS. We present the measurement results using one table for each set of graphs. In each table, besides the correlations with CF and R, we also present the correlation with ANF, as it is the new metric proposed in this paper.

TABLE V:
CORRELATIONS WITH CF, R AND ANF UNDER DDD GRAPHS

|       | CF      | R       | ANF     |
|-------|---------|---------|---------|
| NodeC | -0.8267 | 0.8243  | 0.8267  |
| CF    | 1.0000  | -0.9963 | -1.0000 |
| AC    | -0.8291 | 0.8208  | 0.8291  |
| NatC  | 0.9999  | -0.9960 | -0.9999 |
| NST   | -1.0000 | 0.9963  | 0.9999  |
| EGR   | -0.9998 | 0.9962  | 0.9998  |
| R     | -0.9963 | 1.0000  | 0.9963  |
| ANF   | -1.0000 | 0.9963  | 1.0000  |

Table V gives the eight metrics' correlation results with CF, R, and ANF respectively under the DDD graphs. An important observation from this table is the following: *if a metric is positively correlated with CF, then it is negatively correlated with R, and vice versa*. This shows that, in general, when a graph is robust under random failures, it is fragile under targeted attacks, and vice versa. This can be explained as follows. When the nodes in a graph are similar (e.g., in terms of node degrees), it is difficult to carry out targeted attacks since no nodes have significant importance. Thus, the graph is robust against targeted attacks. On the other hand, when the nodes in a graph have a large degree variance, then typically a small portion of nodes will have very large degrees and the majority of nodes will have small degrees. This will make random failures mostly occur at those nodes with small degrees, so the graph is robust against random failures.

TABLE VI:
CORRELATIONS WITH CF, R AND ANF UNDER BA GRAPHS

|      | CF      | R       | ANF     |
|------|---------|---------|---------|
| NodeC | -0.3540 | 0.2371 | 0.5396 |
| CF   | 1.0000  | -0.5613 | -0.7678 |
| AC   | -0.0970 | 0.0005  | 0.1226  |
| NatC | 0.9546  | -0.4770 | -0.7394 |
| NST  | -0.8581 | 0.7746  | 0.9710  |
| EGR  | -0.6606 | 0.7009  | 0.9542  |
| R    | -0.5613 | 1.0000  | 0.7770  |
| ANF  | -0.7678 | 0.7770  | 1.0000  |

TABLE VIII:
CORRELATIONS WITH CF, R AND ANF UNDER WS GRAPHS

|      | CF      | R       | ANF     |
|------|---------|---------|---------|
| NodeC | -0.3101 | 0.2203 | 0.2785 |
| CF   | 1.0000  | -0.1134 | -0.9872 |
| AC   | 0.4645  | -0.0805 | -0.4609 |
| NatC | -0.2981 | -0.0609 | 0.2595  |
| NST  | -0.0269 | 0.0563  | 0.0573  |
| EGR  | 0.3472  | -0.0083 | -0.3154 |
| R    | -0.1134 | 1.0000  | 0.1049  |
| ANF  | -0.9872 | 0.1049  | 1.0000  |

Other notable observations from Table V include:
- Only NatC strongly and positively correlate with CF, showing that it reflects more on the robustness under random failures.
- NodeC, AC, NST, EGR, and ANF all strongly and positively correlate with $R$, showing that they reflect more on the robustness against targeted attacks.
- NST, EGR, and R all strongly correlate with ANF, showing that these four metrics work similarly to ANF under the DDD graphs.

Table VI gives the eight metrics' correlation results with CF, R, and ANF respectively under the BA graphs. From this table, we can see that all phenomena observed from Table V remain true in the sense of positiveness or negativeness of correlations, and the only differences are that the coefficient values become smaller.

TABLE VII:
CORRELATIONS WITH CF, R AND ANF UNDER ER GRAPHS

|      | CF      | R       | ANF     |
|------|---------|---------|---------|
| NodeC | -0.3747 | 0.1213 | 0.4092 |
| CF   | 1.0000  | -0.4842 | -0.9747 |
| AC   | -0.3562 | 0.1326  | 0.4080  |
| NatC | 0.9525  | -0.3881 | -0.9398 |
| NST  | -0.8732 | 0.4248  | 0.8976  |
| EGR  | -0.7321 | 0.3502  | 0.7612  |
| R    | -0.4842 | 1.0000  | 0.4570  |
| ANF  | -0.9747 | 0.4570  | 1.0000  |

Table VII gives the eight metrics' correlation results with CF, R, and ANF respectively under the ER graphs. Similar to Table VI, this table also shows that all phenomena observed from Table V remain true in the sense of positiveness or negativeness of correlations, and the only differences from Table V are that the coefficient values become smaller.

Table VIII gives the eight metrics' correlation results with CF, R, and ANF respectively under the WS graphs. This table basically agrees with the previous three tables that a metric shows opposite correlation signs with CF and R, except that NatC negatively correlates with both CF and R in this table, though weakly. Another difference from the previous three tables is that NST, EGR, and R no longer strongly correlate with ANF.

*In summary, we can draw the following conclusions from the four tables in this subsection:*

- Based on their tendencies, we can divide the eight metrics into three categories: CF and NatC for random failures; ANF, NST, EGR, and R for targeted attacks; and NodeC and AC weakly for targeted attacks.
- Although the metrics NST, EGR, and R strongly correlate with ANF in the first three tables, they do not in the fourth table. This shows that ANF is not equivalent to these four metrics.
- It is amazing to see that CF has strong correlations (either positive or negative) with NatC, NST, EGR, R, and ANF. This reflects that the reciprocal of CF can actually be used to measure network robustness under targeted attacks. However, CF is not sensitive to network topology change. For instance, a network of a cycle of six nodes will receive the same CF value as another network of two cycles of three nodes, but the latter should be less robust as it is disconnected.

*C. Performance of Graph Models under Eight Robustness Metrics and Four Graph Properties*

In this subsection, we first experiment on how the three popular graph models (BA, ER, and WS) perform on the eight robustness metrics, revealing which graph model achieves the highest value in average given a robustness metric. Then, we experiment to obtain which graph model achieves the highest value in average given a graph property. The latter experiments on graph properties help interpreting the results on robustness metrics. Note that we did not include the set of DDD graphs in the experiments for this subsection, since the set of DDD graphs includes graphs with different kinds of degree distributions, and does not follow a particular graph model.

Table IX gives the average values of the eight robustness metrics on three graph models. Specifically, for each graph model, the average value of a metric on 50 randomly-generated graphs is presented in this table. As discussed in Section III, for EGR, the reciprocals of its values are presented in this table, since the original values of this metric increase when the

TABLE IX:
AVERAGE VALUES OF ROBUSTNESS METRICS ON THREE GRAPH MODELS

|       | BA        | ER        | WS        |
|-------|-----------|-----------|-----------|
| NodeC | 3.5200    | 1.6400    | 4.9200    |
| CF    | 0.9212    | 0.8740    | 0.8608    |
| AC    | 2.0144    | 1.2659    | 0.9462    |
| NatC  | 7.9228    | 4.0302    | 4.2772    |
| NST   | 3.02E+156 | 5.31E+166 | 2.18E+168 |
| EGR   | 0.00013   | 0.00015   | 0.00015   |
| R     | 0.25573   | 0.35358   | 0.38058   |
| ANF   | 5.24582   | 6.46080   | 7.34037   |

network robustness decreases. From this table, we can mainly see that:
- For the metrics NodeC, NST, EGR, R, and ANF, the WS graphs perform the best, i.e., being the most robust among these three graph models.
- For the metrics CF, AC, and NatC, the BA graphs perform the best.

Below we will explain the phenomena seen in this table primarily using what is observed about the graph property DV in the Table X below.

TABLE X:
AVERAGE VALUES OF FOUR GRAPH PROPERTIES ON THREE GRAPH MODELS

|                                     | BA       | ER      | WS      |
|-------------------------------------|----------|---------|---------|
| Degree Variance (DV)                | 45.7636  | 7.5092  | 1.4590  |
| Avg Shortest Path Length (ASPL)     | 2.6118   | 2.7660  | 3.2135  |
| Avg Clustering Coefficient (ACC)    | 0.1092   | 0.0403  | 0.3441  |
| Assortativity Coefficient (AsCo)    | -0.1044  | -0.0152 | -0.0210 |

Table X gives the average values of the four graph properties on three graph models. Similar to the way of obtaining Table IX, this table presents the average value of a property on 50 randomly-generated graphs of a graph model. From this table, we can mainly see that:
- DV exhibits the highest value on the BA model, and the lowest value in the WS model. This explains why WS graphs see higher values on the metrics NST, EGR, R and ANF which favor graphs with small DVs, and why BA graphs see higher values on the metrics CF and NatC which favor graphs with large DVs.
- ASPL exhibits a slightly higher value on the WS model than on the BA and ER models, even though the WS model is for small-world graphs. This is because the BA and ER model also have the small-world feature, meaning that the ASPL of a graph has a logarithmic relationship with $n$. As introduced in [27], the WS model has not only the small-world feature, but also the high clustering coefficient feature which is evident from the ACC row of this table. However, the BA and ER models have more hub nodes (the nodes with large degrees) than the WS model, which gives rise to the smaller ASPL in these two graph models.
- ACC exhibits the highest value on the WS model, which is already explained in the point above.
- AsCo exhibits a value around zero for all three graph models, showing that the graphs in all three models are not assortative.

VII. CONCLUSION AND FUTURE WORK

This paper proposed ANF as a metric for network robustness, and demonstrated that it has the following three advantages. First, since ANF considers network flows among all pairs of nodes in a network, it is sensitive to network topology changes. Second, we have proved that ANF satisfies the 'Strictly Increase' property. Finally, we have given an $O(n^2)$ algorithm for computing ANF by building the Gomory-Hu trees of networks first.

This paper also compared ANF with other seven representative robustness metrics. Different from existing comparison works, our comparisons focus on the scenarios where network topologies change while preserving the same $n$ and $m$. This focus allows us to reveal phenomena not reported before. Important such phenomena include:
- The graph property DV impacts network robustness much stronger than the other three graph properties experimented in this paper: ASPL, ACC, and AsCo.
- The metric NatC strongly aligns with CF for measuring network robustness under random failures; and the metrics ANF, NST, and EGR strongly align with R for measuring network robustness under targeted attacks; and the metrics NodeC and AC weakly align with R for measuring network robustness under targeted attacks.

Moreover, the experimental data reveal that there does not exist a single metric that beats all others in measuring network robustness. Instead, each metric reflects network robustness from a different perspective. ANF, per se, strongly reflects network robustness under targeted attacks, and also embodies the network capability for carrying traffic since it is based on network flows. Thus, this paper is not intended to show that ANF is superior to other metrics, but to add a new angle in measuring network robustness and add a new metric possessing those three advantages aforementioned.

For future work, it is an interesting challenge to find more efficient algorithms for computing ANF. This can be explored in two directions. First, we currently enumerate the network flows among all node pairs by utilizing Gomory-Hu trees and then calculate the ANF by summing them up. An obvious idea is to examine whether the sum can be calculated without enumerating all-pairs network flows. Although this idea is obvious and intriguing, an algorithm with this idea has not been discovered in the literature yet. Second, we currently assume unit capacity on network links to achieve the $O(n^2)$ complexity of calculating ANF. It will be significant to design an algorithm that achieves the same complexity on arbitrary link capacities. At least, no researchers have proved that this is impossible yet.

ACKNOWLEDGMENT

This work is supported by a pilot research grant P00025091 from NSW Cyber Security Network, Australia.